## ***Magnetoresistance in granular $CrO_2$ : influence of crystallographic and magnetic microstructure***

*A. Bajpai and A. K. Nigam*

*Department of Condensed Matter Physics and Material Science*

*Tata Institute of Fundamental Research*

*Homi Bhabha Road, Mumbai-400005*

## Abstract:

We report magnetotransport measurements on high purity sintered samples of spintronic $CrO_2$ in an unexplored crystallographic regime between 5-300 K. The negative magnetoresistance (MR) as derived from RH isotherms is observed to be unhysteretic up to temperatures as high as 200 K. Between 240-290 K, RH isotherms exhibit some unusual features including a positive MR and strong pinning effects. These feature disappear above 290 K and is apparently related with the antiferromagnetic ordering of the insulating grain boundary. Qualitatively similar features with significantly enhanced MR are also observed when the GB density is increased. These results bring out the role played by the magnetic and crystallographic microstructure on the magnitude, sign and hysteresis of the magnetoresistance in this technologically important material.

$CrO_2$ is a well known half metallic[1] ferromagnet (FM) having Curie temperature (Tc) ~ 400 K. It exhibits almost perfect spin polarization[2] (SP) of charge carriers at the Fermi level and thus is a potential candidate for the upcoming area of spintronics[2-8]. In granular form, the electronic conduction between the metallic FM grains of $CrO_2$ takes place through the process of tunneling of charge carriers across the insulating grain boundaries (GB) which appear as a thin layer of thickness 1-2 nm just outside the $CrO_2$ grain[3-8]. These GBs have been identified[4-8] to be an antiferromagnetic (AFM) insulator, $Cr_2O_3$ with a bulk AFM $T_N$ ~307 K[9]. The network of FM grains separated by an insulating GB resemble a typical magnetic tunnel junction (MTJ) having FM electrodes separated by a thin layer of insulator. This makes granular $CrO_2$ a good magnetoresistive material in which the negative MR $[(R_0-R_H)/R_0]$ is consistent with tunneling MR seen in a MTJ, the magnitude of which can be further enhanced by modulations of GB thickness[4,6].

On the flip side, there exist certain practical problems associated with this material, primarily related with its metastable character and difficulties in synthesis[10], which prevents optimization of the MR by modulations in microstructural parameters. Due to this reason the magnetotransport on granular $CrO_2$ have mostly been reported on the commercially available powders with needle like grains of the order of 0.1-0.4 microns[4-6]. These powders exhibit saturation magnetization Ms ~110 emu/g and coercivity Hc ~1000 Oe. The GB density in these samples is usually enhanced either by annealing these powders in air/oxygen[6] or by mixing and cold pressing $CrO_2$ with $Cr_2O_3$ powders[4]. However it is observed that though %MR is large at lower temperatures, it falls rapidly with increasing temperatures and is of the order of ~ 0.1 % near the room temperatures as derived from the extrapolation of low temperature MR data[4]. In this context, it is interesting to note that for an important spintronic material like polycrystalline powders of $CrO_2$, there is lack of experimental data on RH isotherms in the vicinity of room temperatures.

In this letter we report magnetotransport measurements in the temperature range of 5-300 K on $CrO_2$ samples synthesized using a new route[11] in which we could vary the microstructural parameter while retaining the phase purity of $CrO_2$. These samples were characterized using X ray diffraction, SEM and bulk magnetization measurements[12]. The samples are in the form of sintered pellets having grains in the form of micron size rods (average length ~ 5 micron) and Ms ~127 emu/g. The MH isotherm at 5K exhibit a very small remanence ($M_R$~5-10 emu/g)) and coercivity ( $H_c$ ~ 50 Oe) which slightly increase near room temperature[12]. Significantly, the %MR as derived from RH isotherms do not show any hysteresis till about 200 K and is also considerably large near the room temperatures. Certain new features could be unmasked, particularly between 240-300 K which were not seen in for granular $CrO_2$ prior to this work.

The magneto transport data were taken using PPMS (model Quantum Design PPMS-9) using standard 4 probe technique in transverse geometry. The RH isotherm for pure $CrO_2$ at 5K is shown in Fig. 1a, exhibiting usual negative MR as has been observed previously but with negligible hysteresis. In commercial powders the Hc ~1000 Oe and resistance usually increases with applied field below Hc and shows considerable hystersis at lower fields[4]. Unlike previous reports[4,7], the unhystertic pattern in RH isotherms persists right up to 200 K and is consistent with the observation of considerably small $H_C$ and $M_R$ as seen in our samples. Thus below 200 K the scattering associated with domain walls intrinsic to the $CrO_2$ grain or even the magnetic nature of GB (which can introduce irreversible pinning centers through the localized states) is negligible as there are no history effects and hence it is crystallographic processes that presumably dominate the MR patterns. (Inset of Fig. 1a). Moreover, the enhanced grain size plays a crucial role, leading to significant reduction in coercivity which is related to the observation of unhystertic MR till about 200 K.

Several unusual features were observed in RH isotherms recorded above 200 K. Fig. 1b shows a representative graph at 250 K. Here (i) R was found to increase with H, thereby producing a positive MR, in the forward field cycle (below 1 Tesla) and a distinct slope change was seen above the saturation field (~ 1 Tesla for $CrO_2$). (ii) The RH isotherms in this temperature range exhibit strong pinning effects and residual resistance after first field cycling is larger then the zero field resistance. These data imply that magnetic history effects become prominent above 200 K and arise not only from the temperature variation of magnetization, Hc and $M_r$ intrinsic to the FM grain, but also its coupling to the AFM GB. The AFM character of the tunnel barrier is likely to introduce its own temperature and field dependence and can also inhibit the free switching of the magnetization of FM.

It is to be recalled that the observation of positive MR (where the resistance increases with increase in applied field) has been observed in other magnetic tunnel junctions[13-15] including the tunnel junction consisting of Co-$CrO_2$[14]. Here the sign of MR and consequently the SP is seen to be influenced by the choice of barrier material[13], the electronic structure at the interface and the bias voltage[14-15] etc. In our case, the observation of positive MR between 240–290 K appears to arise from the FM/AFM interface and is substantiated from measurements on RH isotherms above 300 K. At these temperatures, the AFM ordering associated with the GB weakens and the %MR remains negative throughout the field cycle as can be seen in Fig 1c. The observation of residual resistance after first field cycling being larger than the zero field resistance between 240-290 K appears to be related with the magnetic character of GB. Whether these strong pinning effects are intrinsically related to the grain ( shape or stress anisotropy associated with CrO2 grain) or the GB is difficult to attribute from bulk transport/ magnetization measurements alone. However all these unusual features between 240-290 K could be reproduced in samples made in different batches. Fig 1c displays RH isotherm at 300 K for pure $CrO_2$ synthesized in different batches exhibiting negative MR of about 1 %. In Fig. 2 we present RH isotherm recorded on a sample in which the GB density was substantially enhanced by diluting it with $Cr_2O_3$ . This sample has Ms ~70 emu/g,, Hc~50 Oe and exhibits a narrow MH loop, similar to what is seen for pure $CrO_2$. The RH isotherms on this highly diluted sample also exhibit similar qualitative features but the %MR thus derived is significantly enhanced (about 3 %) especially near the room temperature. This is significant considering the fact that though the enhancement of GB density is known to result in improving the MR properties at low temperatures, its impact on the MR near room temperature has been found to be negligible[4].

In conclusion, from the measurements of magneto transport, it is apparent that that there exist various contributions to MR which become prominent at different temperatures. Enhancing the grain size results in significant reduction in Hc and Mr resulting in the observation of unhysteretic MR till about 200K. The AFM nature of GB plays a dominant role above 200 K which appears to be related to the positive MR in forward cycle. The hysteresis in field cycling include contributions not only with the intrinsic FM nature of grains but also from pinning of the spins at FM/AFM interface. The magneto transport data clearly bring out the role of AFM nature of GB near the room temperature and suggests that nature of interface of the FM/AFM phase, the electronic structure of the band and its temperature evolution needs to be investigated by detailed band structure studies in these compounds.

**Acknowledgement:** Authors thank Prof. S.K Malik for extending facilities in his lab during the course of this work and Prof. S. Bhattacharya for many helpful discussions.

# Figure Captions

**Figure 1**

**(a)** RH isotherms at 5K measured on pure $CrO_2$. This sample contains grains in the form of micron size rods exhibiting Ms~127 emu/g. Inset shows RH isotherms for the same sample exhibiting negative and unhysteretic MR at various temperatures **(b)** RH at 250 K for the same sample. The forward cycle exhibits a positive MR and strong irreversibility together with pinning effects. **(c)** RH isotherm (circles) for the same sample at 300 K exhibiting negative MR~1%. Data as obtained on a sample of pure $CrO_2$ prepared in a different batch is shown for the sake of comparison (stars) .

**Figure 2.** RH isotherms as measured on granular $CrO_2$ in which the GB density is substantially enhanced by intentionally diluting $CrO_2$ with about 50 % of $Cr_2O_3$. Inset shows positive MR ( ~3 %) in the forward cycle and strong irreversibility effects on the same sample.

**Fig. 1:**

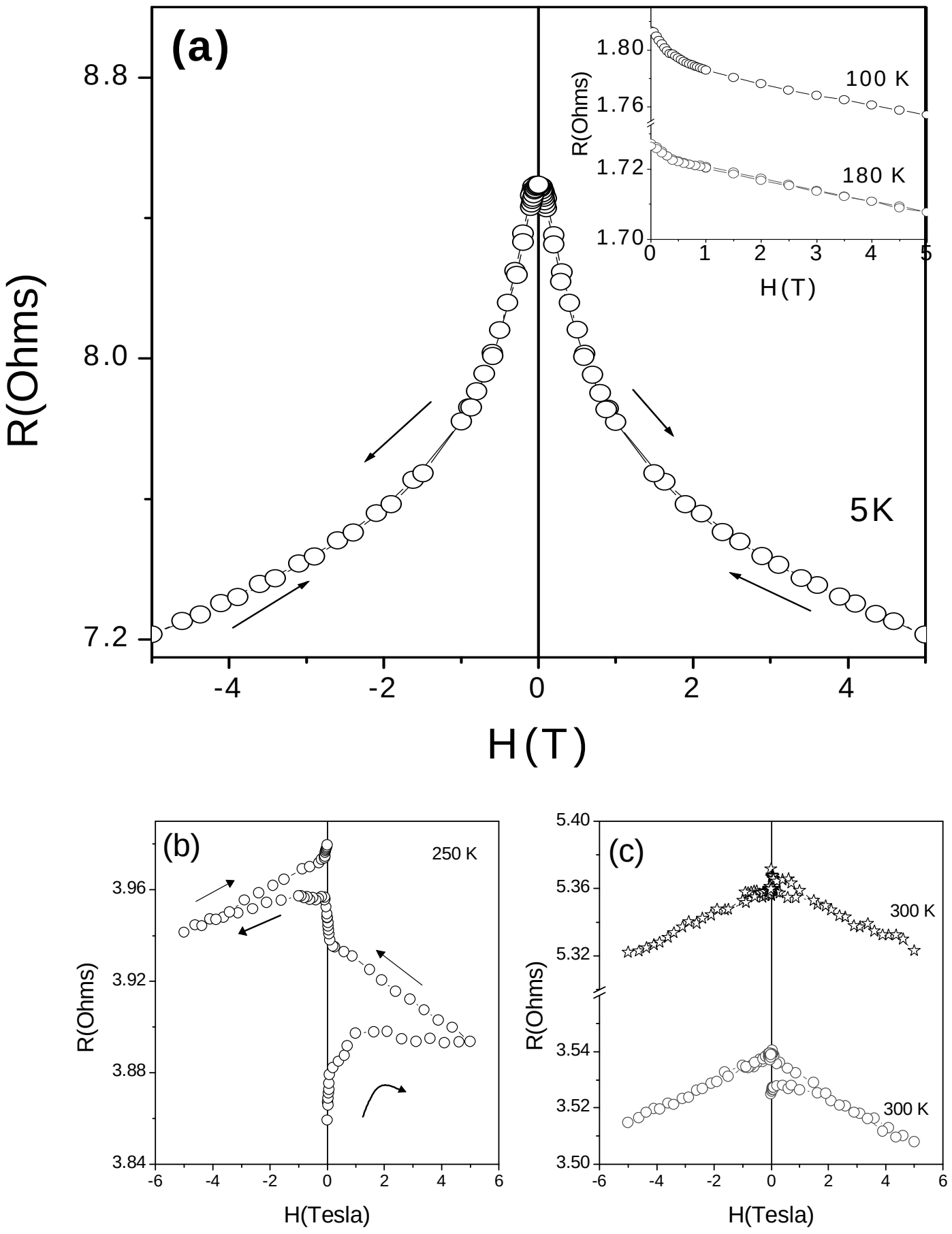

**Fig 2:**

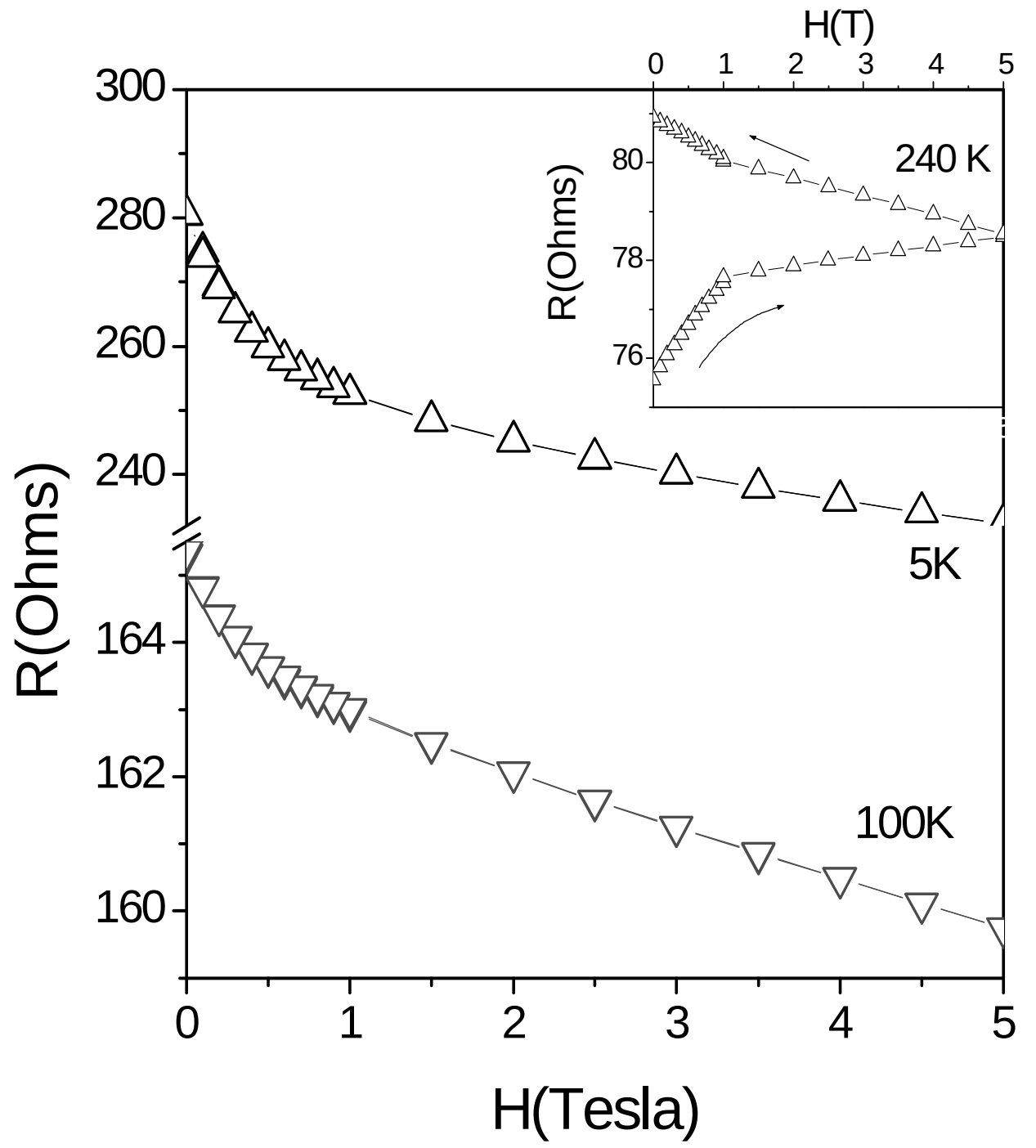